\documentstyle[preprint,revtex]{aps}
\begin{document}
\draft
\begin{title}
Semiclassical Effects and the Onset of Inflation
\end{title}
\author{Esteban Calzetta}\cite{ec}
\begin{instit}
Instituto de Astronom\'\i a y F\'\i sica del Espacio

cc 67, suc 28, (1428) Buenos Aires, Argentina
\end{instit}
\author{Maria Sakellariadou}\cite{ms,pa}
\begin{instit}
Research Group on General Relativity

Facult\'e des Sciences, Universit\'e Libre de Bruxelles

CP 231 Campus Plaine, Boulevard du Triomphe, 1050 Bruxelles, Belgium
\end{instit}
\receipt{}
\begin{abstract}
We present a class of exact solutions to the constraint
equations of General Relativity coupled to a Klein - Gordon
field, these solutions being isotropic but not homogeneous.
We analyze the subsequent evolution of the consistent Cauchy
data represented by those solutions, showing that only certain
special initial conditions eventually lead to successfull
Inflationary cosmologies. We argue, however, that these
initial conditions are precisely the likely outcomes of
quantum events occurred before the inflationary era.
\end{abstract}
\section{Introduction}
\label{Intro}
In this paper we address the issue of
the relevance of semiclassical
cosmological effects to Inflation in
the Very Early Universe. To this end, we consider
spherically symmetric models in which the geometry is
coupled to a massive, free, minimally coupled real
scalar field. We propose a particular choice of gauge,
which allows us to solve Einstein's constraint equations
( see below ) exactly, in closed form, in the case of
interest. We then use the self
consistent Cauchy data so obtained to discuss the
necessary conditions for ``natural'' Inflation (see
below), and match these conditions against those
likely to result out of a semiclassical era.
{}From this analysis, we conclude that semiclassical
effects enhance the likelihood
of Inflation.

The combination of the Hot Big Bang Model and Inflation
provides at present the most comprehensive picture of the
evolution of the Universe at our disposal. To the well
known successfull quantitative predictions of the
Big Bang model ( such as the existence and temperature
of the Cosmic Microwave Background, the relative
abundances of light elements and the red shift of
distant galaxies\cite{cosmo} ), Inflation adds a plausible
explanation
for the near critical value of the observed density in the
Universe\cite{turner}.
It also predicts a
primordial fluctuations spectrum  whose amplitude has the right
order of magnitude, as checked against recent observations\cite{cobe}
However, this class of cosmological models
rests on the assumption of highly specific
initial conditions for the cosmic evolution. This situation
is hardly acceptable, specially because
alleviating the fine tuning of initial conditions, as
required by the Big Bang model, was historically one of the
motivations to investigate inflationary cosmologies\cite{guth}.

It should be stressed that the proper study of the
necessary conditions for Inflation requires the consideration
of nonlinear effects; indeed, linear perturbations of
an Inflationary solution are eventually redshifted away,
and cannot affect the global characther of the cosmic evolution\cite{linear}.
However, nonlinear perturbations can decouple themselves
from the Hubble flow and destroy the homogeneity that
Inflation is supposed to bring about.

Non linear perturbations of an Inflationary model have
been considered by a number of authors\cite{authors}, most notably in
a series of numerical simulations by Goldwirth and Piran\cite{dalia}.
These analysis concur in that Inflation demands initial
conditions which are already homogeneous over regions
larger than the original horizon size. This homogeneity
cannot be accounted for through ( classical ) physical
processes acting in the pre - Inflationary era.
Therefore, to achieve successfull Inflation in
classical cosmology, some
fine tuning of initial
conditions cannot be avoided.

The details of how initial conditions should be fixed
vary with the different versions of the inflationary model in the
literature. In this paper, we shall consider only the ``chaotic''
inflationary model\cite{linde}, where Inflation is sustained by the
stress energy of a free Klein Gordon field, in an Universe
born from an initial singularity. We choose these models
because their extreme simplicity makes them the most
universal of inflationary cosmologies.

The limitations on initial conditions required by chaotic inflation
can be understood in a back of the envelope calculation,
as follows. Suppose  an inflaton field $\phi$ of
mass $m$, varying over distances $\lambda$, being the
source for an inflationary expansion with Hubble constant
$H\sim m\phi /m_p$, where $m_p$ stands for Planck's mass
. The condition of vacuum dominance
requires $m\lambda\gg 1$ and $\dot\phi\ll m\phi$, where a
dot is a derivative with respect to cosmological time. On the
other hand, under conditions of slow roll  over, $\dot\phi\sim
m^2\phi /H$, from where we get $\phi\gg m_p$, and
$\lambda\gg m^{-1}\gg H^{-1}$. A
more carefull analysis shows that $\lambda$ should be some ten
times the horizon size\cite{noialtri}.

While this result shows that Inflation in no way frees the
present state of the Universe from dependence on initial
conditions, the question remains about the relative
likelihood of inflationary initial conditions as opposed
to generic ones. Nevertheless, here too, as long as we remain
within the framework of classical cosmology, Inflation
proves itself of little use. General Relativity being a
time reversal invariant theory, final
and initial configurations are in a one to one correspondence.
Therefore no mechanism such as Inflation can enhance the likelihood
of a particular set of final states
( of course, this holds as long as one does not introduce an
ad hoc measure in configuration space ). In other words, the
inflationary hypothesis does not render a homogeneous and
isotropic Universe any more likely than simply to assume
homogeneity and isotropy to begin with.

The situation changes when semiclassical effects are taken into
account. Indeed, while quantum fields in curved spaces
evolve unitarily, it is generally impossible in concrete
situations to determine exactly the quantum state of a given
field.   For  example,  while  we  know  the  occupation  numbers  of the
different modes in the Background Microwave Radiation, we would
be hardly pressed to determine the correlations between different modes
as well. Therefore, in practice, quantum fluctuations act as a
bath or environment for the macroscopic degrees of freedom of the
Universe, and these evolve following a dissipative effective
dynamics, whereby time reversal invariance is
broken\cite{timebreak}. Under these circumstances,
 while we know that the present state of the
Universe cannot be rendered totally independent of initial conditions,
it makes sense again to ask for the relative likelihood of
inflating cosmologies.

A  similar set  of  questions  have  already  been  investigated  in  the
framework of homogeneous cosmological models. In ref.\cite{IPCCA} it has been
shown that the back reaction of conformal particles created by
a Taub Universe helps to delay recollapse\cite{wald}, and thereby increases
the likelihood of Inflation ( in these models, Inflation occurs
when the Universe
reaches a size such that the cosmological constant
overpowers shear in the Hamiltonian constraint ). This was done by
following the evolution self consistently, in the approximation
were the back reaction of created particles is taken into account,
but vacuum polarization terms are neglected.

In this paper, we shall study the relevance of semiclassical
effects to Inflation in inhomogeneous but isotropic, with
respect to a singled out point, models. In this context, to
solve for the evolution self consistently is no longer possible.
Rather, we shall follow the strategy of concentrating on a
particular Cauchy surface, which is assumed to lie at the
future of the semiclassical era, but at the past of the
inflationary one. Thus, the Cauchy data on this surface are
themselves the outcome of the semiclassical era. We shall
study self consistent sets of Cauchy data, that is, solutions
of Einstein's constraint equations, and obtain conditions for
a set of data to lead to ``natural''Inflation. We shall
consider Inflation to be natural if there is a significative
probability that, well after Inflation has begun, any
nondescript observer will find him or herself into an
Inflating, near Friedmann - Robertson - Walker like region. Then we shall
show that these conditions are already contained in the requirement
of consistence with an earlier semiclassical era.
In other words, we shall conclude
 that semiclassical effects select Cauchy data
leading to natural Inflation.

Our analysis shall proceed in the Hamiltonian, or ADM\cite{MTW},
 formulation
of General Relativity. Any spherically symmetric metric comprises
two physical degrees of freedom, and two Lagrange multipliers,
the lapse function, and the shift in the radial direction; the
Klein - Gordon field introduces an extra degree of freedom. The
Lagrange multipliers are associated to two nontrivial constraints,
which in turn allow us to impose two arbitrary gauge conditions on
the model ( we shall use this freedom to simplify the equations below,
rather than opting for a gauge invariant formulation ). Our tactic
shall be to use the constraints and the gauge freedom to fix
entirely the space metric and the geometrodynamical momenta; the
lapse and shift are then defined by the requirement of consistency
of the dynamical Einstein's equations. This approach leaves the
Klein - Gordon equation ( written in Hamiltonian form ) as the only
dynamical law.

In this paper we shall keep a specific physical situation in mind.
We found it convenient to take advantage of the gauge freedom in
General Relativity to obtain this particular scenario in its simplest
form. For this reason, we shall adopt a ``custom made'' gauge,
rather than any of the usual choices, such as maximal slicing\cite{MTW},
a Tolman - like metric\cite{landau}, or a synchronous gauge.
For the same reason, we shall develop an analysis of the Einstein -
Klein - Gordon system from first principles, rather than apply the
general solutions available in the literature\cite{BCMN}.

The situation of interest is as follows.
 As we have seen
in our ``back of the envelope'' argument above, successfull
chaotic Inflation requires a very high and homogeneous initial
value of the field ( we shall not discuss here the possible sources
of such field values; they occur in quantum cosmological
models based on the Hartle - Hawking `` no boundary '' proposal\cite{qc}).
Whatever the mechanism to provide such
configuration, it is physically likely that quantum and/or
thermal fluctuations will result in creating ``holes'' in the
field, that is, regions where the value of the field gets closer
to its ( vanishing ) equilibrium value. Of course a deep enough
``hole'' will not inflate at all, but a shallow ``hole'' may be
capable of inflating, thus becoming an Inflationary island or bubble
in a larger Universe. This island is, nevertheless, surrounded
by a transition layer where conditions are far
from homogeneity. The naturalness condition, over and above the
usual conditions on the field for there
to be Inflation, concerns the relative sizes of the island and the
transition layer.

For simplicity, we shall consider a single island,
so we shall assume that the field is high and homogeneous far from
the origin. Under these conditions, the metric will approach
asymptotically a Friedmann - Robertson - Walker ( FRW ) form.
We shall use our gauge
freedom to force the three metric to keep a FRW
form everywhere. Departures from homogeneity shall be coded
into the lapse and shift functions. As we shall see below,
in this gauge we shall be able to display the dependence of
these functions on the field and its conjugated momentum in
detail. This, in turn, will allow us
 to translate  the naturalness condition into a
concrete inequality for the Cauchy data. In a number of cases
of interest, such as when the field momentum vanishes, the
field itself being arbitrary, we shall obtain
 closed form, exact expressions
for lapse and shift. This is already a vast
generalization over previously known results\cite{authors,dalia,noialtri};
 the general case
can be handled perturbatively.

Allthough we shall not discuss the evolution of these
Cauchy data in detail, we shall show below that the naturalness
condition puts limits on the shift function accross the
transition layer. On the other hand, as we shall discuss
in more detail in the body of the paper, we expect the
Universe to be conformally flat ( vanishing Weyl tensor )
by the end of the quantum era\cite{bianchii}. The exact form of the
self consistent Cauchy data previously obtained shall allow us
to show that, in this model, conformal flatness results in
the same type of conditions than naturalness. In this sense,
therefore, it can be said that semiclassical effects select
for naturally inflating Cauchy data.

This result is both
of relevance to cosmology, and of great interest as a
non trivial application of quantum field theory in curved spaces
(QFTCS). This theory being only an approximation to a yet
unknown fully quantized theory of gravity, its meaningful
applications are confined to weakly quantum effects,
where gravitational quantum behavior is not expected to be
relevant. Given these restrictions, QFTCS is only capable
to yield entirely new results in phenomena where quantum
effects are able to acummulate over time, there being
no classical phenomena in a position to screen them. The
canonical example where these conditions are fullfilled is
Black Hole evaporation\cite{hawking}, which is still now possibly the main
area of development in QFTCS. Conformal particle creation,
from the vacuum, in cosmological settings, is similarly a
cumulative, intrinsically quantum phenomenon.
The study of the effect of particle creation processes on
the shape of our Universe is therefore, beyond its
relevance to cosmology, one of the few areas where QFTCS
leads to observable predictions, not to be obtained in
classical theory.

The paper is organized as follows. In next section we present
the model and the solution of Einstein's constraint equations.
In section III, we discuss the conditions for natural Inflation,
the effect on Cauchy data of a semiclassical era, and the
relationship between the two. In section IV we briefly state our
conclusions. A technical discussion of the solution of the
constraint equations for nonvanishing field momenta is included as
an appendix.
\section{Self Consistent Spherically Symmetric Cauchy Data}
\label{Selfco}
\subsection{Canonical Variables and Hamiltonian}
\label{Cano}
In this section, we shall carry out an analysis of spherically
symmetric solutions of the constraint equations of the Einstein -
Klein - Gordon system, in order to discuss in the next section which
Cauchy data eventually lead to acceptable inflationary cosmologies.
We shall adopt for our discussion the ADM formalism, whose
starting point is the $3+1$ decomposition of the space - time metric
as
\begin{equation}
ds^2=-N^2dt^2+g_{ij}(dx^i+N^idt)(dx^j+N^jdt)\label{3+1}
\end{equation}
Here, $g_{ij}$ is the metric induced on a space like surface, on which
the $3+1$ decomposition is based, and $N$, $N^i$ are the lapse and
shift functions which describe the embedding of the spatial surface
into the four dimensional space time ( we follow MTW
\cite{MTW} conventions; latin
indexes run from $1$ to $3$ ). Assuming spherical symmetry, we
may simplify Eq.(\ref{3+1}) to
\begin{equation}
ds^2=-N^2(R,t)dt^2+A^2(R,t)(dR+\nu (R,t)dt)^2+B^2(R,t)d\Omega\label{3+1ss},
\end{equation}
where we introduced $\nu = N^1$ and $d\Omega =d\theta^2
+\sin^2\theta d\varphi^2$. It is convenient
to choose the conformal degree of freedom of the space metric
as one of the geometrodynamical variables.
Therefore, we rewrite Eq. (\ref{3+1ss}) as
\begin{equation}
ds^2=-N^2(R,t)dt^2+e^{2\alpha (R,t)}
\{ X^{-4/3}(R,t)(dR+\nu (R,t)dt)^2+X^{2/3}(R,t)d\Omega\}\label{3+1ssaX}
\end{equation}
The peculiar parametrization of the conformal metric will be of use
below.

If we take $\alpha$ and $X$ as geometrodynamical canonical coordinates,
then the canonical momenta shall be parametrized in terms of two
independent degrees of freedom, $\Pi_{\alpha}$ and $\Pi_X$,
canonically conjugated to the respective variables. The full
momentum tensor density, in $R,\theta ,\varphi$ coordinates,
becomes
\begin{equation}
\Pi^i_j=({1\over 6})[\Pi_{\alpha}\delta^i_j+{3\over 2}X\Pi_X
(\delta^i_j-3\delta^i_1\delta^1_j)]\label{momenta}
\end{equation}
With this parametrization, the ``kinetic'' term in the Hamiltonian
\begin{equation}
K=({16\pi\over m_p^2})Ng^{-1/2}\{\Pi^i_j\Pi^j_i-(1/2)(\Pi^i_i)^2\}\label{kin}
\end{equation}
($m_p$ is Planck's mass )becomes
\begin{equation}
K=({16\pi\over m_p^2})
{Ne^{-3\alpha}\over 24}\{ -\Pi^2_{\alpha}+9X^2\Pi_X^2\}\label{kinaX}
\end{equation}
The ``shift'' part of the Hamiltonian
\begin{equation}
S=\Pi^{ij}(N_{i;j}+N_{j;i})\label{shift}
\end{equation}
becomes
\begin{equation}
S=(1/3)\Pi_{\alpha}(\nu '+3\alpha '\nu )-X\Pi_X\nu '+X'\Pi_X\nu ,
\label{shiftaX}
\end{equation}
where a prime means a derivative with respect to $R$.
Finally, the ``potential'' term
\begin{equation}
V=(-{m_p^2\over 16\pi})Ng^{1/2}{\cal R}\label{pot}
\end{equation}
( where ${\cal R}$ is the spatial curvature )
is best computed by observing that the spatial metric is conformally
flat. Indeed, introducing a new radial coordinate $r$ through
\begin{equation}
{dR\over X}={dr\over r}\label{Rtor}
\end{equation}
The spatial metric becomes
$e^{2\omega}[dr^2+r^2d\Omega ]$, where
\begin{equation}
\omega =\alpha +(1/3)\ln X -\ln r\label{omega}
\end{equation}
Therefore, we find
\begin{equation}
{\cal R}=(-1)e^{-2\omega}\{ 4r^{-2}\partial_rr^2\partial_r\omega +
2 (\partial_r\omega )^2\}\label{spaceR}
\end{equation}
or, in terms of $R$ derivatives
\begin{equation}
V=({m_p^2\over 16\pi}) Ne^{\alpha}X^{4/3}
[4\alpha ''+2\alpha '^2+
({16\over 3})\alpha '({X'\over X})+({4\over 3})({X'\over X})'+
({14\over 9})({X'\over X})^2
-2X^{-2}]\label{potaX}
\end{equation}
The matter field introduces a new canonical variable $\phi$ and
its conjugated momentum $\Pi_{\phi}$, and a new term in the
Hamiltonian
\begin{equation}
H_m=\nu \phi '\Pi_{\phi}+{N\over 2}\{ e^{-3\alpha}\Pi_{\phi}^2
+e^{\alpha}X^{4/3}\phi '^2+e^{3\alpha}m^2\phi^2\}\label{Hmat}
\end{equation}
where $m$ is the mass of the minimally coupled, real, non
interacting field $\phi$. The total Hamiltonian is given by
$K+S+V+H_m$.
\subsection{Field Equations and Gauge Conditions}
\label{fieldeq}
Having found the Hamiltonian of the model in the previous
section, the field equations, in Hamiltonian form, are found
by taking variational derivatives in the usual way. Variation
with respect to lapse and shift yields the Hamiltonian and
Momentum constraints
\begin{eqnarray}
&&~({16\pi\over m_p^2})
{e^{-3\alpha}\over 24}\{ -\Pi^2_{\alpha}+9X^2\Pi_X^2\}\nonumber\\
&&+({m_p^2\over 16\pi}) e^{\alpha}X^{4/3}
\{4\alpha ''+2\alpha '^2+
({16\over 3})\alpha '({X'\over X})+({4\over 3})({X'\over X})'+
({14\over 9})({X'\over X})^2
-2X^{-2}\}\nonumber\\
&&+{1\over 2}\{ e^{-3\alpha}\Pi_{\phi}^2
+e^{\alpha}X^{4/3}\phi '^2+e^{3\alpha}m^2\phi^2\}=0\label{Hcon}
\end{eqnarray}
\begin{equation}
(-1/3)\Pi_{\alpha}'+\alpha '\Pi_{\alpha}+(X\Pi_X)'+X'\Pi_X+\phi
'\Pi_{\phi}=0\label{Mcon}
\end{equation}
Variation with respect to the momenta yields the velocities
\begin{equation}
\dot\alpha ={1\over 3}(\nu '+3\nu\alpha ')-({16\pi\over m_p^2})
({Ne^{-3\alpha}\over 12})\Pi_{\alpha}\label{alphadot}
\end{equation}
\begin{equation}
\dot X=-X\nu '+X'\nu+({16\pi\over m_p^2})
({3Ne^{-3\alpha}\over 4})X^2\Pi_X\label{Xdot}
\end{equation}
\begin{equation}
\dot\phi =Ne^{-3\alpha}\Pi_{\phi}+\nu\phi '\label{phidot}
\end{equation}
( where a dot stands for time derivative ).
Variation with respect to $\phi$ yields
\begin{equation}
\dot\Pi_{\phi}=(\nu\Pi_{\phi})'+(Ne^{\alpha}X^{4/3}\phi ')'-
Ne^{3\alpha}m^2\phi ,\label{piphidot}
\end{equation}
which, toghether with Eq.\ (\ref{phidot}), is equivalent to the
Klein - Gordon equation. It is unnecessary to take variations
with respect to $ \alpha $ and $ X $, as the resulting equations
are dependent on those already derived.

We are thus left with six equations for eight unknowns, which
must be supplemented with two arbitrary gauge conditions. As
discussed in the Introduction, we envisage solutions which
approach Friedmann - Robertson - Walker ( FRW) behavior at infinity,
where the field shall be assumed to be homogeneous. This
boundary condition shall be easiest to implement in a gauge
where the three metric is constrained to be already in FRW
form everywhere, so that deviations from homogeneity are
encoded solely in the lapse and shift functions. Therefore
we impose as gauge conditions
\begin{equation}
\alpha '=0\label{1gauge}
\end{equation}
\begin{equation}
X=3R\label{2gauge}
\end{equation}
The extreme simplicity  of the functional dependence of $X$ is
the reason behind our unconventional parametrization of the
space metric ( the metric becomes explicitly FRW under the
change of coordinates $R=r^3/3$ )
. These gauge conditions still allow for time
redefinitions; the gauge can be totally fixed by demanding,
e. g. , that the lapse function approaches $1$ as $R\to\infty$.

The field equations acquire a simpler form in terms of the
non canonical variables $P_{\alpha}=e^{-3\alpha}\Pi_{\alpha}$,
$P_{X}=e^{-3\alpha}\Pi_{X}$, and $P_{\phi}=e^{-3\alpha}\Pi_{\phi}$.
The constraints become
\begin{eqnarray}
&&~({2\pi\over 3m_p^2})
\{ -P^2_{\alpha}+9(3R)^2P_X^2\}\nonumber\\
&&+{1\over 2}\{ P_{\phi}^2+m^2\phi^2
+e^{-2\alpha}(3R)^{4/3}\phi '^2\}=0\label{gaugeHcon}
\end{eqnarray}
\begin{equation}
(-1/3)P_{\alpha}'+
3RP_X'+6P_X+\phi 'P_{\phi}=0\label{gaugeMcon}
\end{equation}
The dynamical equations for $\alpha$ and $X$ now become consistency
conditions for the lapse and shift, namely
\begin{equation}
\dot\alpha ={1\over 3}\nu '-({4\pi\over 3m_p^2})
NP_{\alpha}\label{galphadot}
\end{equation}
\begin{equation}
\nu '-{\nu\over R}=({36\pi\over m_p^2})RNP_X
\label{gXdot}
\end{equation}
The Klein - Gordon equation takes the form
\begin{equation}
\dot\phi -\nu\phi ' =NP_{\phi}\label{gphidot}
\end{equation}
\begin{equation}
\dot{(e^{3\alpha}P_{\phi})}=e^{3\alpha}(\nu P_{\phi})'
+e^{\alpha}(N(3R)^{4/3}\phi ')'-
Ne^{3\alpha}m^2\phi\label{gpiphidot}
\end{equation}
\subsection{Solving the constraint equations}
\label{solvcons}
We turn now to the study of the solutions of the constraint
and consistency equations derived in the previous subsection.
To this end, it is convenient to introduce the notation
\begin{equation}
H^2=({4\pi\over 3m_p^2})[P_{\phi}^2+m^2\phi^2+e^{-2\alpha}(3R)^{4/3}
\phi '^2]\label{h2}
\end{equation}
In the homogeneous case, $H$ reduces to the Hubble constant.
We also parametrize $P_{\alpha}$ and $P_X$ in terms of a new
variable $\xi$, as follows
\begin{equation}
P_{\alpha}=-({3m_p^2\over 4\pi})H\cosh\xi\label{palpha}
\end{equation}
\begin{equation}
P_X=-({3m_p^2\over 4\pi}){H\over 9R}\sinh\xi ,\label{pX}
\end{equation}
whereby the Hamiltonian constraint is reduced to an identity,
and the momentum constraint becomes
\begin{equation}
\xi '+({1\over 2R})(e^{2\xi}-1)={H'\over H}+({4\pi\over m_p^2}){\phi 'P_{\phi}
e^{\xi}\over H}\label{xieq}
\end{equation}
If the last term in Eq.\ (\ref{xieq}) could be neglected, the
general solution would be
\begin{equation}
e^{2\xi}\sim{H^2\over{\cal H}^2+{f\over R}}\label{feq}
\end{equation}
where $f$ is a constant, and
\begin{equation}
{\cal H}^2={1\over R}\int_0^R~dR'~H^2(R')\label{calh2}
\end{equation}
(${\cal H}^2$ is therefore an smoothed out version of $H^2$ ). Indeed,
we find
\begin{equation}
{\cal H}^2{}'={H^2-{\cal H}^2\over R}\label{calh2prime}
\end{equation}
\begin{equation}
\xi '\sim{H'\over H}-({1\over 2R})[{H^2-({\cal H}^2+{f\over R})
\over {\cal H}^2+{f\over R}}],\label{xiprime}
\end{equation}
from where it is easy to verify Eq.\ (\ref{xieq}). In the general case,
we uphold the ansatz Eq.\ (\ref{feq}), but now allowing $f$ to be a
function of $R$ and $t$; substituting into Eq.\ (\ref{xieq}), we
find
\begin{equation}
f'=-({8\pi\over m_p^2})R\phi 'P_{\phi}\sqrt{{\cal H}^2+{f\over R}}\label{fpr}
\end{equation}
To avoid a singularity at the origin, we must adopt the boundary
condition $f(0)=0$. Therefore, in the case in which $P_{\phi}$
vanishes on the initial surface ( but the field profile is
arbitrary ), or if the field is homogeneous, no matter the form of
$P_{\phi}$, $f$ is identically zero, and our solution of the
constraints is exact.

We shall return to the perturbative solutions to Eq.\ (\ref{fpr})
in the Appendix.
\subsection{Solving for Lapse and Shift}
\label{solapse}
Having reduced the constraints to the single Eq.\ (\ref{fpr}) for
the unknown function $f$, we now turn to consider the consistency
conditions Eqs.\ (\ref{galphadot}), (\ref{gXdot}). We first
notice they imply the simple relationship between lapse and shift
\begin{equation}
\nu=3R\{\dot\alpha-NHe^{-\xi}\}\label{shilap}
\end{equation}
Replacing Eq.\ (\ref{shilap}) into Eq.\ (\ref{gXdot}), we obtain
\begin{equation}
R{(NHe^{-\xi})'\over (NHe^{-\xi})}={1\over 2}(e^{2\xi}-1)\label{intern}
\end{equation}
Now, Eqs.\ (\ref{feq}) and (\ref{calh2prime}) can be combined to
yield
\begin{equation}
e^{2\xi}-1=R\{(\ln [{\cal H}^2+{f\over R}])'-{f'\over R
({\cal H}^2+{f\over R})}\}\label{2intern}
\end{equation}
And so, using again Eqs.\ (\ref{feq}) and
(\ref{fpr}), and imposing the boundary
condition that the lapse should approach $1$ as $R\to\infty$, we find
\begin{equation}
N=e^{\{-({4\pi\over m_p^2})\int_R^{\infty}~dR'[{\phi 'P_{\phi}
\over\sqrt{{\cal H}^2+({f\over R})}}](R')\}}\label{bigN}
\end{equation}
\begin{equation}
\nu=3R\dot\alpha\{1-({\sqrt{{\cal H}^2+{f\over R}}\over \dot\alpha})
{}~e^{(-({4\pi\over m_p^2})\int_R^{\infty}~dR'[{\phi 'P_{\phi}
\over\sqrt{{\cal H}^2+({f\over R})}}](R'))}\}\label{bignu}
\end{equation}
These expressions, toghether with the gauge chosen form of
$\alpha$ and $X$, completely determine the space time metric,
once the field configuration is given. Allthough they are not yet
in closed form, since the function $f$ is known only implicitly,
as the solution of Eq.\ (\ref{fpr}), they are explicit enough
to serve our purposes.
In the next section, we shall
apply them to study the conditions under which natural
Inflation occurs in the Universe.
\section{Natural Inflation and Semiclassical Effects}
\label{Natinf}
\subsection{Conditions for Natural Inflation}
\label{Conatinf}
In the previous section, we analyzed the Hamiltonian structure of
spherically symmetric Einstein - Klein - Gordon systems, in a
manner adapted to the further discussion of Inflationary
cosmology, but keeping nevertheless full generality.  In this
subsection, we shall introduce a number of new assumptions,
which will allow us to specialize the general formalism to
an specific physical situation, a nonlinear perturbation
of an otherwise successful chaotic inflationary model. We
shall derive from this analysis the conditions under which
Inflation may still be obtained after the perturbation.

Concretely, we have in mind a situation where, in an homogeneous
cosmological model, a bubble  is created, either through
quantum or thermal tunneling, where the field is lower than
its spatial average value. Since most  bubbles are created
in an instantaneously stationary configuration\cite{coleman},
 we shall assume that
$P_{\phi}$ vanishes  in a neighborhood of the bubble. In the
exterior region, notwithstanding, slow roll over conditions
hold, and $P_{\phi}\sim -(3\dot\alpha )^{-1}Nm^2\phi$. The
field is homogeneous both within and outside the bubble\cite{coleman};
 for
concreteness, we shall assume $\phi '\ge 0$ throughout.

Let us first show how the expected result that inflation shall proceed
both in and out the bubble, but at different rates, is recovered
form the analysis of the previous section. In the exterior
region, using the slow roll over value for $P_{\phi}$ in
Eq.\ (\ref{fpr}), we obtain
\begin{equation}
f'\sim ({8\pi\over 3m_p^2\dot\alpha})
RNm^2\phi\phi '\sqrt{{\cal H}^2+{f\over R}}~~,\label{fprsro}
\end{equation}
which, in view of the definition of $H^2$, Eq.\ (\ref{h2}), can be
reduced to
\begin{equation}
f'\sim ({RNH^2{}'\over\dot\alpha})~\sqrt{{\cal H}^2+{f\over R}}\label{fprred}
\end{equation}
The solution of this equation is
\begin{equation}
f\sim R(H^2-{\cal H}^2)\label{ansatz}
\end{equation}
Indeed, for this form of $f$, we obtain $\xi =0$ ( cfr.  Eq.\ (\ref{feq})),
$P_X=0$ ( cfr. Eq.\ (\ref{pX})), and
$P_{\alpha}=-({3m_p^2\over 4\pi})H$. From the consistency
conditions Eqs.\ (\ref{galphadot}) and (\ref{gXdot}), and the
requirement that Inflation obtains at infinity, we get
$N\sim \dot\alpha /H$ and $\nu\sim 0$. Eq.\ (\ref{fprred}) reduces to
$f'=R(H^2)'$, which is easily seen to follow from Eqs.\ (\ref{ansatz})
and (\ref{calh2prime}). In regions where $H^2$ is essentially a constant,
moreover, the cosmological time $\tau$ is related to coordinate
time $t$ through $dt/d\tau = 1/N= H/\dot\alpha$, which shows
that $H$ is indeed the ``local'' Hubble constant, as measured by
a comoving observer.

Let us consider now a neighborhood of the bubble. Since $P_{\phi}\sim 0$
here, we have $f\sim 0$ ( because of Eq.\ (\ref{fpr}) and the boundary
condition $f(t,0)=0$ ). Also, from Eq.\ (\ref{bigN}), we see that
$N\sim~{\rm constant}~\sim\dot\alpha/H_{out}$, where $H_{out}$ is the
value of $H$ just outside the transition layer. Inside the bubble
proper, the field is homogeneous, and therefore ${\cal H}\sim H\sim
{\rm constant}~\equiv H_{in}$. Eq.\ (\ref{bignu}) now gives
\begin{equation}
\nu =3R\dot\alpha\{ 1-{H_{in}\over H_{out}}\}\label{bubblenu}
\end{equation}
The full metric reads
\begin{equation}
ds^2=-({\dot\alpha\over H_{out}})^2dt^2+e^{2\alpha }
\{ (3R)^{-4/3}(dR+3R\dot\alpha\{ 1-{H_{in}\over H_{out}}\}
dt)^2+(3R)^{2/3}d\Omega\}\label{bubble3+1}
\end{equation}
Let us define a new radial variable
\begin{equation}
\zeta =Re^{3\int~\dot\alpha\{ 1-{H_{in}\over H_{out}}\}~dt'}\label{zeta}
\end{equation}
In terms of $\zeta$, the metric becomes
\begin{equation}
ds^2=-({\dot\alpha\over H_{out}})^2dt^2+
e^{2\int~\dot\alpha\{{H_{in}\over H_{out}}\}~dt'}
\{ (3\zeta )^{-4/3}(d\zeta )^2+(3\zeta )^{2/3}d\Omega\}\label{rebubble}
\end{equation}
Which takes explicit FRW form under the further changes
$d\tau =({\dot\alpha\over H_{out}})~dt$, $\zeta =(1/3)\rho^3$. We
see that, as expected, Inflation is obtained inside the bubble, with
$H_{in}$ as the Hubble constant measured by comoving observers.

We conclude that, for successful Inflation, the usual conditions
of high and homogeneous field must hold inside the bubble.
Moreover, if the bubble is several horizons in size originally,
it cannot recollapse entirely, since the wall of the bubble
cannot exceed the speed of light, but the Hubble flow is not so
limited. An observer deep inside the bubble, therefore, will
be allowed to assume that he or she is living in a FRW Universe,
irrespective of the conditions outside the bubble.

Conditions in and outside the bubble are linked through the
requirement  of naturalness, however. In the simplified
model we have considered, we obtained two inflationary
regions, the interior and exterior of the bubble, separated
by a transition layer. In a more realistic model, we would
consider a Universe composed of many bubbles, each with its
own surrounding wall. Naturalness is the requirement that,
at any generic instant, the volume inside the bubbles
should be a significative fraction of the volume inside
the walls ( indeed, if we were to carry the Copernican
principle to extremes, we should demand that the volume
within the bubbles be much larger than the volume outside
them ). A similar condition, easier to
implement, is that each
bubble should be comparable in size to the wall surrounding
it.

Let  us apply the later version of naturalness to our model.
As an estimate of the size of the wall at any given moment,
we shall use the forward light cone of the original
transition layer. This overestimates the actual wall, but
is nevertheless appropiate to our purposes, because
observers within the light cone can see the wall, and
therefore are aware that they do not live in a FRW cosmology.

To follow the evolution of the light cone it is not
necessary to solve the Klein Gordon equation in any
detail. On the contrary, by continuity, the limit
light rays emerging from the walls shall behave
essentially like the light rays of the Inflationary
regions, and these are essentially slowly evolving
de Sitter geometries. Concretely, the light cone shall
expand a distance of overall $H_{out}^{-1}$, measured
in the $r$ coordinate, into the exterior region,
and of $H_{in}^{-1}$, measured in the $\rho$ coordinate,
into the bubble ( recall that $r$ and $\rho$ are
the coordinates in which the metric takes explicitly
the usual FRW form, outside and inside the bubble,
respectively ).

To compare the relative sizes of bubble and wall, however,
we must use an uniform coordinate system. If we choose
the $r$ coordinate, for example, we see that the line
$\rho\sim\rho_0\equiv{\rm constant}$ marking the
inner boundary of the wall, becomes
\begin{equation}
r =\rho_0e^{-\int~\dot\alpha\{ 1-{H_{in}\over H_{out}}\}~dt'}\label{wall}
\end{equation}
So, for most times, the bubble is actually exponentially
small with respect to the wall ( since we are assuming that
the expansion rate inside the bubble is lower than outside ).
Of course, we do not require the bubble to be comparable
to the wall for all times, but only for the $60$ or so
e - foldings that Inflation lasts. So, in this model,
naturalness means that the exponential contraction of the bubble
should not be too noticeable for the first $60$ e - foldings
of Inflation. The necessary condition for this, as follows
from Eq.\ (\ref{wall}), is
\begin{equation}
\{ 1-{H_{in}\over H_{out}}\}\le{1\over 60}\label{cond}
\end{equation}
Only Cauchy data satisfying this condition shall lead to
natural Inflation, even if they otherwise fulfill the
requirements for enough Inflation in and out the bubble,
separately. From the point of view of classical
cosmology, however, naturalness, requiring a correlation
between regions several horizons apart,
can only be imposed through fine tuning of initial conditions.
The possibility remains, however, that semiclassical
effects actually favor initial conditions, for the
classical era, already satisfying Eq.\ (\ref{cond}).
In order to investigate this possibility, we turn now to
a discussion of the relevance of semiclassical effects
in cosmology.
\subsection{Semiclassical Effects and Natural Inflation}
\label{seminat}
As discussed in the Introduction, our goal is to determine
whether the consideration of semiclassical cosmological
effects turns naturally inflating Cauchy data any more
likely than simply to assume them {\it ad hoc}. For
this to be the case, it is necessary that semiclassical
effects should break the one to one correspondence
between the cosmic states before and after the
semiclassical era ( or, more exactly, that semiclassical
effects should invalidate Liouville's Theorem as applied
to the Universe ). It follows from this that semiclassical
effects resulting merely in the replacement of the Einstein -
Hilbert action by a more complicated, but still real and local,
effective action, will not affect the likelihood of
natural Inflation and can be disregarded.

The situation changes, however, when
we consider particle creation processes\cite{parker}.
 Indeed, under suitable
statistical assumptions, particle creation defines an arrow
of time\cite{timebreak,henry}.
More important to our present concern, the back reaction of
the created particles has a smoothing effect on the evolution,
thus making certain field configurations ( those leading
to no creation ) prefered above others\cite{parker,lenz}.
 Thus, to quote
a well known example, the possibility of conformal
particle creation makes an isotropic universe a
prefered alternative to anisotropic ones. For simple
cosmological models, such as Bianchi type I universes\cite{bianchii},
it is actually possible to follow in detail the
process of particle creation and isotropization. Moreover,
it has been shown that the connection between these
phenomena is not limited to the semiclassical era,
extending as well to the fully quantum one\cite{timebreak,ADQC}.

The details of how particle creation proceeds in a given
model depend on the peculiar matter content of the model. In
the absense of a generally accepted theory of elementary
particle physics, there is no absolute criteria for
what a realistic theory should look like. However, as far as
we are mostly concerned with processes occurring early
on in the semiclassical era, we can make some simplifications.
Indeed, the main effects of
a strong gravitational field on elementary particles can
be described by allowing masses and
coupling constants to run according to their renormalization group
equations, the scale being fixed by suitable curvature
invariants\cite{david,milwaukee}.
It follows that asymptotically free theories of elementary
particles actually become free in the early Universe, and that
masses can be ignored. For spin $1/2$ and $1$, minimally coupled
fields, this implies they become approximately conformally invariant.

For scalar fields the situation is slightly more complex, because,
while conformal coupling is a fixed point of the 1 loop
Renormalization Group equations, higher loop corrections
tend to make it unstable\cite{david}. These small deviations from conformal
invariance are not important, nevertheless, because, in any case,
creation of scalar particles is less efficient than that of
gauge and spinor fields.

The creation rate for conformal particles in nearly conformally
flat Universes is given by $NC^2/80\pi$
, where $C^2$ is the square
of the Weyl tensor, and $N$ the effective number of  particle
species, defined as $N=N_1+N_{1/2}/4+N_0/24$, $N_i$ being the
number of species of spin $i$\cite{bianchii}.
 The value of $N$ depends upon
the particular theory of elementary particles being used; for
typical GUTs, $N\sim 100~{\rm to}~1000$\cite{ross}.

In view of the earlier discussion, any initial configuration
with a nonzero Weyl tensor will be brought towards
conformal flatness by particle creation; correspondingly, towards
the end of the quantum era, the Weyl tensor should be
negligible against the conformal part of the curvature. In
a near inflationary evolution, the curvature scale is set by
the Hubble constant, and our argument shows that the
inequality $C^2H^{-4}\ll 1$ must hold.

To compare this constraint on physically acceptable Cauchy
data to the condition for natural Inflation, Eq.\ (\ref{cond})
above, we must compute the Weyl tensor for the specific form
of the metric corresponding to a neighborhood of the bubble.
Assuming slow roll over conditions, that is, that space
derivatives dominate over time derivatives, except for
the conformal factor, we obtain
\begin{equation}
C^2=12H_{out}^4\{ R\beta '-3R^{2/3}[R^{4/3}\beta\beta ']'\}^2
\sim 12H_{out}^4\{ R\beta '\}^2,\label{weyl}
\end{equation}
where
\begin{equation}
\beta =\{ 1-{{\cal H}\over H_{out}}\}\label{beta}
\end{equation}
{}From Eq.\ (\ref{calh2prime}) above, it is easy to see that
\begin{equation}
R\beta '=-{H^2-{\cal H}^2\over 2H_{out}{\cal H}}\label{rbeta}
\end{equation}
Recall now  that $H_{out}$ is the limit value of $H$ as we
approach the edge of the transition layer. There is then a region,
close to the outer rim of the layer, where $H\sim H_{out}$.
In this region
\begin{equation}
R\beta '=-\beta [{1-\beta /2\over 1-\beta}]\sim -\beta\label{aprbeta}
\end{equation}
and the condition of smallness on the Weyl tensor translates
into $\beta\ll 1$. But recall also that, from its definition,
${\cal H}$ is slowly varying compared to $H$ itself. Therefore,
across the transition layer, even if already $H\sim H_{out}$,
we still have ${\cal H}\sim H_{in}$, and $\beta\ll 1$ actually
implies the condition Eq.\ (\ref{cond}) for natural inflation.

To summarize, we have shown that, even if a mechanism were found
to create regions of large, homogeneous inflaton field, classical
Inflation would still be unnatural, because the volume of the
transition
layers among these regions would be typically larger than the
regions themselves. However, semiclassical effects forbid
Cauchy data where the local Hubble constant changes
strongly between inflating regions. Therefore, in semiclassical
cosmology, successful Inflation is also natural, without the
need of extra assumptions correlating the value of the Inflaton
field in the different inflating domains. In this sense,
semiclassical Inflation is more natural than its classical
counterpart.
\section{Conclusions}
The discussion in this paper has progressed in three stages, each
using the results of the previous  one, but still essentially
independent. In the first stage, we analyzed the constraint
equations of General Relativity coupled  to a Klein - Gordon
field. Assuming spherical symmetry, we showed that the general
solution to these constraints could be expressed in terms of a
single unknown function. In many cases of interest, this
function may be determined exactly, yielding consistent
Cauchy data in closed, analytic, form. In the general case,
the unknown function may be computed perturbatively.

This result has been achieved through the use of a special
gauge choice, devised to simplify the constraint equations to
the utmost. The only dynamical law in this approach is the Klein -
Gordon equation itself. The gain in simplicity of the constraints
is paid for in terms of the complexity which this equation acquires.
However, being hyperbolic, the Klein - Gordon
equation is amenable to a qualitative treatment, which discloses the
general features of the cosmological evolution.

In the  second  stage  of  our  reasoning,  we developed such a
qualitative analysis, seeking to determine the necessary
conditions for what we called ``natural'' inflation. Of
course, to obtain Inflation at all certain special initial
conditions must be assumed, involving very high and
homogeneous initial values of the inflaton field. Our
discussion aimed to show that, even if these
conditions were assumed locally, the resulting Universe
would still be very different to an Inflationary one, unless
one further condition were added, ``naturalness'', linking the
values of the field over several horizon lenghts. This
result is not dependent upon the full details of
the Cauchy data previously found; however, knowledge of
those data allowed us to translate naturalness into a
concrete inequality, which the Cauchy data must
satisfy, in order to lead to an admissible cosmology.

Finally, in the third stage of our argument we confronted
our conditions for ``natural'' Inflation against
the present understanding of semiclassical cosmology. Based
on the well proven smoothing effect of particle creation,
we argued that the Universe should have left the semiclassical
era in a state of near conformal flatness. This already puts
a bound on the possible gradients of the metric elements
at the beginning of Inflation; our explicit form for
those metric elements allowed us to show that this bound actually
implies the naturalness condition previously derived.

There is an important caveat which goes with this argument,
and which we would like to make explicit here. In discussing
the likely evolution of the Universe during the semiclassical
era, we are already assuming that the initial conditions at
its beginning were not too extreme. Indeed, the detailed
models of semiclassical cosmologies in the literature
assume for the most part near Friedmann - Robertson - Walker
conditions\cite{hu}, and it would be unjustified to
extrapolate these results to arbitrarily strong
inhomogeneities. Moreover, it should not be expected that
semiclassical effects could allways stabilize a
classically growing perturbative mode.

However, even under the most conservative reading of the
literature, it should be accepted that, under the statistical
assumptions discussed in the Introduction, semiclassical evolution is
irreversible, and, in particular, that a nontrivial set
of initial conditions is actually brought to conformal
flatness through particle creation and back reaction.
These results are enough to support our main conclusion,
which is that semiclassical effects enhance the likelihood
of Inflation in the Early Universe. This conclusion, in
turn, confirms the findings of previous studies of
inflation in homogeneous models\cite{IPCCA}.

It is certainly likely  that a conclusion along these lines,
given these hypothesis, could have been reached through
general arguments, independent of the detailed form
of the Cauchy data. However, the particular strategy we
have followed is relevant in that it points the way for
further studies of classical and semiclassical Inflation.
For example, the self consistent Cauchy data for the
Einstein - Klein - Gordon system we present here, provide
also an environment where such questions as the
dependence on initial conditions of the spectrum of
density fluctuations in Inflation can be investigated.
The relevant feature of the solutions presented here is, of
course, that in no way we have assumed small departures from
homogeneity.

We are continuing our research on these manyfold questions.
\nonum
\section{Acknowledgements}
We would like to acknowledge the multiple comments and
suggestions from the participants to the NATO Workshop
on Origin of Structure in the Universe ( Pont d'Oye,
Belgium, 1992), specially L. Grishchuk, A. Starobinsky,
W. Unruh and R. Wald, where a preliminary draft of this work
was presented.

E. C. wishes to thank the hospitality of the RGGR group
at the Free University ( Brussels ); M. S. that of the
GTCRG group at IAFE ( Buenos Aires ).

This work was supported by CONICET, UBA and Fundaci\'on
Antorchas ( Argentina ), and by the Directorate General for
Science Research and Development of the Comission of the European
Communities under contract N$^o$ CI1-0540-M(TT).
\nonum
\section{Appendix: Perturbative solutions to the constraints}

In the main body of the paper we have shown that the full geometry
can be parametrized in terms of a single function $f$, obeying
Eq.\ (\ref{fpr}). In particular  cases, such as when $P_{\phi}$
vanishes identically, $f\equiv 0$, and the metric can be
worked out explicitly. However, as we pointed out above, it
would be unrealistic to assume such conditions throughout
space. For successfull inflationary models, moreover, we
must have $f\sim R(H^2-{\cal H}^2)$ for large $R$. It is
interesting, therefore, to investigate the solutions to
Eq.\ (\ref{fpr}) in non trivial cases.

If we assume, as in the main body of the paper, $\phi '\ge 0$
and $P_{\phi}\le 0$ ( in Inflation, $\phi '=0$ and
$P_{\phi}\sim -m^2\phi/3H$ ), then $f'$ is positive, and
$f$ is a non decreasing and non negative function of $R$.
Under these conditions, it is possible to build a sequence
of functions approximating $f$ as follows: the first function
$f_0$ in the sequence is identically zero, and the n-th
is the solution to
\begin{equation}
f_n{}'=-({8\pi\over m_p^2})R\phi 'P_{\phi}\sqrt{{\cal H}^2
+{f_{n-1}\over R}}\label{apfpr}
\end{equation}
With boundary condition $f_n(0)=0$.

We observe that all functions in the sequence are positive and
non decreasing. The sequence itself is nondecreasing at every point:
$f_1$ is certainly larger than zero, because it starts at
zero, and has a positive derivative; $f_2$ is larger than $f_1$,
because both are positive, nondecreasing functions, with
the same value at the origin, and $f_2$ has the larger derivative, etc.
Therefore, the sequence $f_n(R)$ has a limit for every $R$, and the
limit function satisfies Eq.\ (\ref{fpr}).

This method of successive approximations is appealing not only
because convergence is guaranteed, but also because at every step
it is possible to estimate its accuracy. Indeed, from Eqs.\ (\ref{fpr})
and (\ref{apfpr}) we see that
\begin{equation}
(f-f_n)'=-({8\pi\over m_p^2})R\phi 'P_{\phi}\{\sqrt{{\cal H}^2+{f\over R}}
-\sqrt{{\cal H}^2+{f_{n-1}\over R}}\}\label{2apfpr}
\end{equation}
 Since the square root is a convex function, it follows that
\begin{equation}
(f-f_n)'\le -({8\pi\over m_p^2})\phi 'P_{\phi}
\{2\sqrt{{\cal H}^2+{f_{n-1}\over R}}\}^{-1}(f-f_{n-1})\label{3apfpr}
\end{equation}
And a fortiori
\begin{equation}
(f-f_n)'\le h(R) (f-f_{n-1})\label{4ap}
\end{equation}
where
\begin{equation}
h(R)=-({4\pi\over m_p^2}){\phi 'P_{\phi}\over{\cal H}}\label{5apfpr}
\end{equation}
Eq.\ (\ref{4ap})  can be rewritten as
\begin{equation}
(f-f_n)'-h(R) (f-f_n)\le h(R) (f_n-f_{n-1})\label{6ap}
\end{equation}
which in turn leads to
\begin{equation}
f\le f_n+\int_0^R~dR'~h(R')e^{\int_{R'}^R~dR''~h(R'')}(f_n-f_{n-1})
\label{7ap}
\end{equation}
This is the sought for bound on $f$

\end{document}